\documentclass{iaus}
\usepackage{graphicx}

\newcommand\kms{{\rm\,km\,s^{-1}}}

\newcommand\ulsr{U_{\rm LSR}}
\newcommand\vlsr{V_{\rm LSR}}
\newcommand\wlsr{W_{\rm LSR}}
\newcommand\uw{(U^2+W^2)^{1/2}}

\title[The Galactic thin and thick discs in the context of galaxy formation] %% give here short title %%
{The Galactic thin and thick discs in the context of galaxy formation}

\author[Thomas Bensby \& Sofia Feltzing]   %% give here short author list %%
{Thomas Bensby$^1$ \& Sofia Feltzing$^2$}%

\affiliation{$^1$European Southern Observatory, Santiago, Chile; 
email: {\tt tbensby@eso.org}\\[\affilskip]
$^2$Lund Observatory, Lund, Sweden;
email: {\tt sofia@astro.lu.se}}

\pubyear{2009}
\volume{265}  %% insert here IAU Symposium No.
\pagerange{000--000}
\date{?? and in revised form ??}
\jname{Chemical abundances in the Universe -- Connecting first stars and planets}
\editors{Katia Cunha, Monique Spite \& Beatrice Barbuy , eds.}
\begin{document}

\maketitle

\begin{abstract}
We have obtained high-resolution spectra and carried out a 
detailed elemental abundance analysis for a new sample of
899 F and G dwarf stars in the Solar neighbourhood. The results allow
us to, in a multi-dimensional space consisting of stellar ages, 
detailed elemental abundances, and full kinematic information for the 
stars, study and trace their respective origins. Here we briefly 
address selection criteria and discuss how to define a thick disc star. 
The results are discussed in the context of galaxy formation.
\end{abstract}

\firstsection % if your document starts with a section,

%=======================================================================
\section{Introduction}

The study of the Milky Way is important in the context of galaxy formation. 
The hierarchical build-up of galaxies within $\Lambda$CDM, the 
currently most successful theory for the formation of large-scale 
structure in the Universe (e.g., Springel et al.~2006), implies 
that a galaxy like the Milky Way has suffered a bombardment of 
merging blocks over its life time. 
Stellar discs are very fragile to changes in the
gravitational potential, hence the
existence of spiral galaxies like the Milky Way is a challenge for 
$\Lambda$CDM. However, recent studies by, 
e.g., Koda et al.~(2009), Stewart et al.~(2009), and 
Scannapieco et al.~(2009), show that these effects might not be as 
severe as previously envisioned. 

Evidence of the presence of a Galactic {\it thick} disc was first 
presented by Gilmore \& Reid~(1983), showing that the density of stars 
as a function of distance from the Galactic plane towards the Galactic 
north pole could not be explained  by  a single exponential, but rather 
two exponentials with different scale heights and with different 
number densities in 
the plane. Subsequent observational 
studies have established that, (compared to the thin disc) the thick 
disc stellar population is a kinematically hotter population, it rotates
more slowly, it is mainly an old stellar population (as old as the stellar
halo), and that it has different chemical properties, particularly
in the sense that it possesses higher $\rm [\alpha/Fe]$ ratios at
a given metallicity (e.g., Bensby et al.~2003, 2005, 2007b; 
Reddy et al.~2006; Fuhrmann2008).

In the context of galaxy formation and as tests of models
of galaxy formation, it is important to establish the properties
of a disc system of the type observed in the Milky Way.
Therefore, to fully understand why the Milky Way has two different disc
populations, how they  formed, how they evolved, and clarify their
relationships to the stellar halo and the Galactic bulge we have
carried out an extensive study of long-lived F and G dwarf stars 
in the Solar neighbourhood.

%=======================================================================
\section{A new stellar sample}

%-----------------------------------------------------------------
\begin{figure}
\resizebox{\hsize}{!}{
\includegraphics[bb=0 160 640 540,clip]{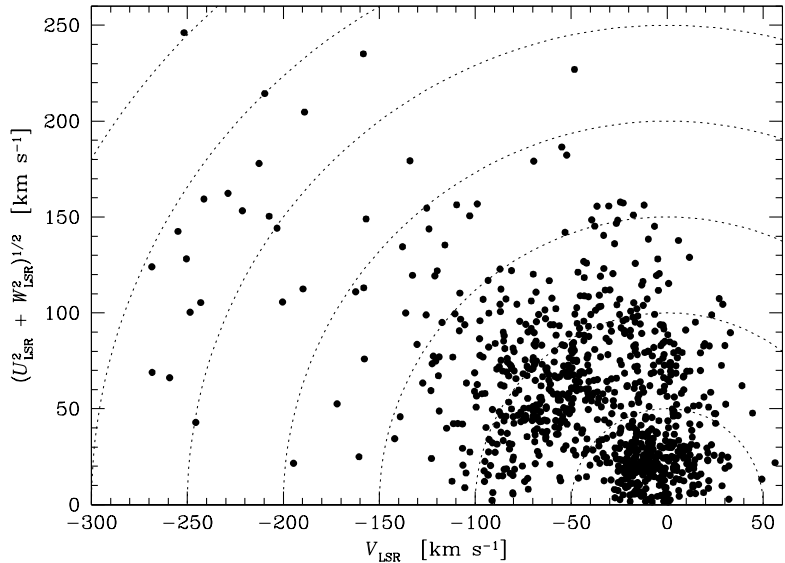}}
\caption{
\label{fig:toomre}
Toomre diagram.
Dotted lines show constant values of the
total space velocity, $v_{tot}=\sqrt{\ulsr^2 + \vlsr^2 + \wlsr^2}$,
in steps of 50\,$\kms$.
}
\end{figure}
%------------------------------------------------------------

We have a sample of 899 nearby F and G dwarf stars for which we have
obtained high-resolution and high signal-to-noise spectra with several
spectrographs (UVES, FEROS, SOFIN, and MIKE; see Bensby et al.~2010, 
in prep. for details).  Among the things we
targeted  during our 
observations are: (i) the oxygen trends in the metal-rich
thin disc (Bensby et al.~2004); (ii) chemical differences between 
the thin and thick discs (Bensby et al.~2003, 2005);
(iii) metal-rich and old stars; (iv) the metal-rich limit of the 
thick disc (first results in Bensby et al.~2007b); (v) the origin
of the Hercules 
stream (first results in Bensby et al.~2007a); (vi) the metal-poor 
limit of the thin disc; (vii) the metal-poor limit of the thick disc; 
(viii) the stars with intermediate kinematics that can not be classified 
as neither thin disc nor thick disk stars (using current 
kinematical criteria, see 
Bensby et al.~2003, 2005); (ix) the metal-rich halo and its interface to the 
thick disc.  Compared to our previous published thin and thick disc
stellar sample that contained 102 dwarf stars (Bensby et al.~2003, 2005),
the current sample is a factor $\sim$8 larger, and will allow us to
put firmer observational constraints on the Galactic discs.

A Toomre diagram, which is a representation of the stars' combined
vertical and radial kinetic energies as a function of the stars' 
rotational energy is shown in Fig.~\ref{fig:toomre}. Low-velocity stars, 
constrained within
$v_{\rm tot}\equiv (\ulsr^{2} + \vlsr^{2} + \wlsr^{2})^{1/2} \lesssim
50\,\kms$
are to a first approximation mainly thin disc stars, and stars with
$v_{\rm tot}$ greater than $\sim70\,\kms$, but less than  
$\sim200\,\kms$, are likely to be thick disc stars (e.g., Nissen~2004). 
Stars with higher velocities are halo stars. The Toomre diagram
shows that our study sample the regions occupied by the thin and 
thick disc stellar populations well. There is also an excess of stars 
with velocities
around $\vlsr\sim-50\,\kms$ and $\uw\sim50-100\,\kms$, the velocity 
space occupied by the Hercules stream (e.g., Bensby et al.~2007a).

%=======================================================================
\section{The dichotomy of the Galactic disc}

%-----------------------------------------------------------------
\begin{figure}
\resizebox{\hsize}{!}{
\includegraphics[bb=0 160 620 695,clip]{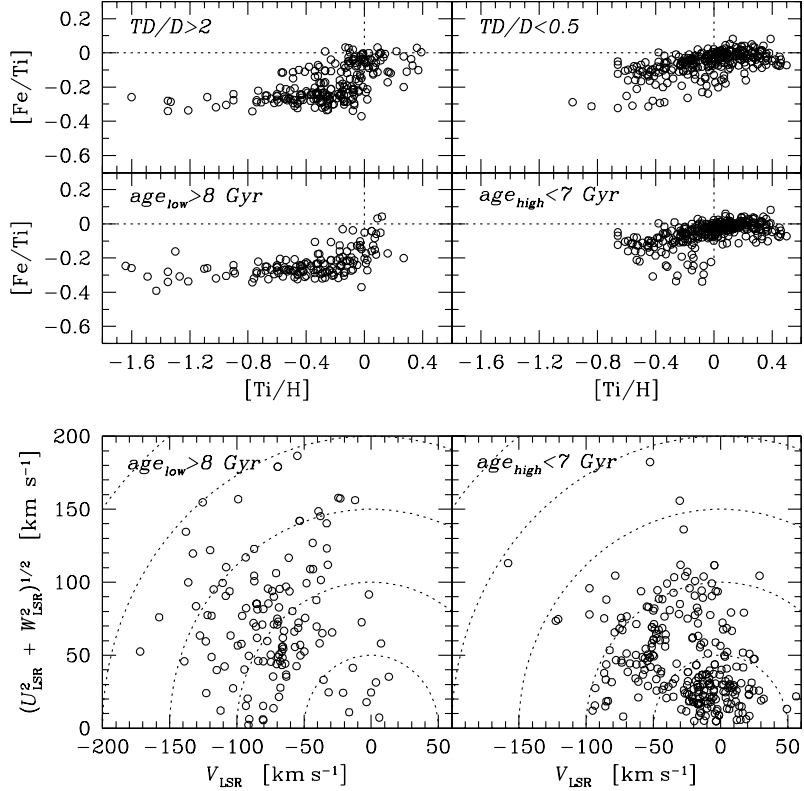}}
\caption{
\label{fig:ageortdd}
Thin disc (right-hand side) and thick disc (left-hand side)
abundance trends using two different selection criteria. 
The stars in the top panels have been selected using the
kinematical criteria used in Bensby et al.~(2003, 2005),
i.e., that a star has to be at least two times likely to belong
to a stellar population than to any of the other populations.
In the middle two panels the stars have been divided 
according to their ages: one sample with lower age estimates
greater than 8\,Gyr, and one sample with upper age estimates
less than 7\,Gyr. The two bottom panels show the Toomre diagrams
for the old and young samples in the middle panels.
}
\end{figure}
%------------------------------------------------------------

One of the aims with our study is to characterize the thin and
thick discs in terms of detailed elemental abundances and stellar
ages, and from there draw further conclusions about the origin and 
evolution of the two discs. As high-resolution observations of a large
sample of F and G dwarf stars currently limits us to the stars
of the immediate Solar neighbourhood, typically the sphere
covered by the Hipparcos mission, we have to rely on kinematical 
criteria when selecting our stellar sample. The {\it top panels}
of Fig.~2, show the $\rm [Fe/Ti]-[Ti/H]$ abundance trends 
for stars that are at least two times more likely of being thick
disc stars than thin disc stars (top left panel), and vice versa
in the top right panel. It is clear that these two sub-samples,
only being different in terms of their kinematic properties, show quite different abundance trends. It is at the same time also clear that
in each of these plots, there exist a number of stars that deviate 
from the large majority, and that actually would fit
better with the other kinematic sample. This is most likely an
effect of the fact that the velocity ellipsoids of the thin a
thick discs partially overlap. Hence, when using kinematical criteria 
to pick either  thin or thick disc stars we a likely to pick a few
high-velocity thin disc stars in the thick disc sample, and  a few
low-velocity thick disc stars in the thin disc sample.

As most investigations indicate that the thick disc mainly is an
old stellar population and that the thin disc is considerably younger,
it might be illustrative to try and separate the sample by age
criteria instead. This is what we have done in the middle panels
of Fig.~2: one sample of stars whose ages have a lower age estimate
of 8\,Gyr (left panel) and one sample of stars whose ages have an 
upper age estimate lower than 7\,Gyr (right panel). 
Now we see two
abundance trends very alike the ones we obtained by
using kinematical criteria only. However, the number of ``outliers"
have been reduced and the abundance trends are ``cleaner".
The bottom panels in Fig.~2 show Toomre diagrams for these two
age-selected samples and we see that the old sample mainly is 
a kinematically hot sample. The young sample, on the other hand, 
contains a lot of kinematically hot stars as well,  stars that,
when using kinematical criteria, would be selected as thick disc stars.
This ``kinematic confusion" demonstrates that one has to be
careful when using kinematical criteria to select thin and thick disc 
stars, and that better criteria to separate the two populations are
needed. However, Fig.~2 shows that, and if focus is not put on individual 
stars, kinematic criteria still can be used to obtain 
stellar samples of the Galactic stellar populations in order to trace 
their evolution.

%=======================================================================
\section{Discussion}

The mostly regarded scenario for the formation of the Galactic thick
disc is that it is a result of an ancient merger event between 
the Milky Way and another (dwarf) galaxy 
(see, e.g., Quinn et al.~1993; Villalobos \& Helmi~2009). 
However, recent simulations
by Sch\"onrich \& Binney~(2009) that show that it is possible 
to form a thick disc in the Galaxy, without a merger event, through
radial mixing of gas and stars. Their model is able to reproduce 
all of the current observables (abundances, stellar ages, dichotomy
of the Galactic disc, etc). So, currently, it appears as if the observable
that would serve in favour of any these two formation scenarios is if there
is a hiatus in the star formation history between the two discs. If present,
the merger scenario would be favoured, as the model by 
Sch\"onrich \& Binney~(2009) implies a continuous star formation history.

%=======================================================================
\begin{acknowledgments}

Sofia Feltzing is a Royal Swedish Academy Research Fellow supported by 
a grant from the Knut and Alice Wallenberg Foundation. 

\end{acknowledgments}

%=======================================================================

\end{document}